\documentclass[a4paper,10pt]{article}
\usepackage{amssymb}
\usepackage[dvips]{color}
\usepackage{graphicx}
\usepackage{amsmath}
\usepackage{a4wide}

\newtheorem{theorem}{Theorem}

\newtheorem{definition}[theorem]{Definition}
\newtheorem{example}[theorem]{Example}

\newtheorem{proposition}[theorem]{Proposition}
\newtheorem{remark}[theorem]{Remark}

\newcommand{\F} {\ensuremath{\mathcal{F}}}
\newcommand{\Prob} {\ensuremath{\mathbb{P}}}
\newcommand{\Q} {\ensuremath{\mathbb{Q}}}

\newcommand{\R} {\ensuremath{\mathbb{R}}}

\definecolor{Bleu}{rgb}{0,0,0.5}

\begin{document}

\title{A Goal Programming Model with Satisfaction Function for Risk Management and Optimal Portfolio Diversification}
\author{Marco Maggis\thanks{Department of Economics, Management and Quantitative Methods, University of Milan, Italy. Email: marco.maggis@unimi.it. The author acknowledges the financial support provided by the European Social Fund Grant.} \and Davide La Torre\thanks{Department of Economics, Management and Quantitative Methods, University of Milan, Italy. Email: davide.latorre@unimi.it}}

\maketitle

\begin{abstract}
We extend the classical risk minimization model with scalar risk measures to the general case of set-valued risk measures. The problem we obtain is a set-valued optimization model and we propose a goal programming-based approach with satisfaction function to obtain a solution which represents the best compromise between goals and the achievement levels. Numerical examples are provided to illustrate how the method works in practical situations.
\end{abstract}


\section{Introduction}
Risk measures are real valued functionals defined on a space of random variables which encloses every possible financial position. It may seem naive to use a single number to describe the complexity of the distributions characterizing those random variables. On the other hand this appears as the only way to succeed in the assessment of the capital requirement needed to a bank to recover a high possible loss due to risky investments. Without any doubt it is a critical point the choice of the axioms defining the
risk measures; these have been vividly discussed since the very beginning of this theory, opening a broad new branch of research
which still triggers the interest of the Mathematical Finance world.

\bigskip

Seminal contributions to this topics were surprisingly given by Bernoulli in 1738 \cite{bern}, who perceived the role of the risk aversion in decision making. This was the starting point of the notion of {\em Expected Utility}. Later on, in a
financial environment, many risk procedure were introduced. It was the case of the \emph{Mean-Variance} criterion (Markovitz, 1952,\cite{Mark}), the \emph{Sharpe's ratio} (1964,\cite{Sha}) and the \emph{Value at Risk} ($V@R$),
defined through the quantiles of a given distribution with a predefined level of probability. This last method is the most employed in credit institutes and has been pointed out as the reference parameter by the Basel Committee on Banking Supervision (Basel II 2006).

\bigskip

At the end of the Nineties, Artzner, Delbaen, Eber and Heath produced a rigorous axiomatic formalization of coherent
risk measures, led by normative intent. The regulating agencies asked for computational
methods to estimate the capital requirements, exceeding the unmistakable lacks showed by
the extremely popular $V@R$. Given a vector space of random variables $L$, the definition of a {\em coherent risk measure} $\rho: L\to \R$  requires four main hypotheses to be satisfied: monotonicity, cash additivity, positive homogeneity, sublinearity (see Section \ref{RM}). The relevant role of the axioms was deeply discussed  in many papers: F$\ddot{o}$llmer and Schied (2002,\cite{FoSchA}), Frittelli and Rosazza Gianin (2002,\cite{FrR}) independently studied the convex case weakening positive homogeneity and sublinearity. El Karoui and Ravanelli relaxed the cash additivity axiom to cash subadditivity (2009,\cite{ER09}) when the market presents illiquidity; Maccheroni et al. (2010,\cite{CMMMa}) showed how quasiconvexity describes better than convexity the principle of diversification, whenever cash additivity does not hold.
\\
Finally two important recent generalizations were introduced by Jouini et al. (2004,\cite{JM04}), who defined set-valued coherent risk measures, and by Hamel and Heyde (2010,\cite{HH10})
who introduced the notion of set-valued convex risk measure. This approach is absolutely natural as far as the risk is expressed and hedged in different currencies.

\bigskip

Diversification plays a crucial role in insurance and financial business and the interpretation of this notion was the source of this vivid debate. An agent who considers a fixed basket of financial instruments $X=(X_1,...,X_d)\in L^{\infty}_{d}$  tries to  reallocate his wealth, by means of a diversified strategy $\alpha\in \R^d$, in order to minimize the risk of his portfolio. Namely, given a real valued risk measure (as introduced in \cite{ADEH}) $\rho:L^{\infty}\rightarrow \R$ we have the following optimization model
\begin{equation}\label{RealOptI}\min \left\{\rho\left(\alpha\cdot X \right)\mid \alpha\in\R^d_+: \sum_{i}\alpha_i=1\right\}
\end{equation}
where $\alpha\cdot X$ is the usual scalar product in $\R^d$.
\\The optimal risk allocation is a classical problem in mathematical economics and it is interesting from both practical and theoretical perspectives.
In more recent years this problem has also been studied in many other contexts such as risk exchange, assignment of liabilities to daughter companies, individual hedging problems (see, for instance, the papers by Heath and Ku (2004,\cite{HK04}), Barrieu and El Karoui (2005,\cite{BE05}), Burgert and Ruschendorf (2006,\cite{BR06}), Jouini et al. (2007, \cite{JST07}),
Acciaio (2007,\cite{Ac07})).

\bigskip

The aim of this paper is to provide a computational procedure, based on the Goal Programming (GP) model, to problem (\ref{RealOptI}) if the agent has to find the best compromise among different beliefs which may come either from the uncertainty on the probabilistic model $\Prob$, or from different opinion that the agent has to face in his institution. These multi-criteria will be aggregated in a unique measure of risk which will be described by a set-valued map $R:L_d^{\infty}\rightrightarrows \R^n$. In particular $d$ stands for the number of financial instruments considered, whereas $n$ is the number of different criteria (which in general might be larger that $d$ as shown in Section \ref{examples}). Thus the interpretation we are giving to $R$ appears pretty different to the original one provided in \cite{JM04}.

\bigskip

The paper is organized as follows: in Section \ref{RM} we recall some basic notions on set valued risk measures, Section \ref{optimal} is devoted to the extension of the optimization problem (\ref{RealOptI}) to the case of set-valued risk measures, Section \ref{MOP} introduces a GP model with satisfaction function and finally Section \ref{examples} presents some numerical results.

\section{Risk Measures} \label{RM}

Let $(\Omega,\mathcal{F} ,\mathbb{P} )$ a probability space and
we denote with $L^0=:L^{0}(\Omega ,\mathcal{F},\mathbb{P})$ the space of $%
\mathcal{F}$ measurable random variables that are $\mathbb{P}$ almost surely
finite. We denote by $L^{\infty}=:L^{\infty}(\Omega ,\mathcal{F},\mathbb{P})$ the space of $\Prob$-almost surely bounded random variables which becomes a Banach lattice once endowed with the $\Prob$-almost sure pointwise partial order and the usual norm of the supremum. In this probabilistic framework we recall the definition of risk measure.

\begin{definition} A risk measure is a functional $\rho:L^{\infty}\rightarrow \R$ which satisfies
\begin{itemize}
\item[i)] {\it monotonicity}, i.e. $X_1\leq X_2$ implies $\rho(X_1)\geq \rho(X_2)$ for every $X_1,X_2\in L^{\infty}$,
\end{itemize}
Moreover a risk measure may satisfy
\begin{itemize}
\item[ii)] {\it convexity}, i.e. $\rho(t X_1 + (1-t) X_2) \le t \rho(X_1) + (1-t) \rho(X_2)$ for all $t\in[0,1]$.
\item[iii)] {\it cash additivity}, i.e. $\rho(X+c)=\rho(X)-c$,
\item[iv)] {\it positive homogeneity}, i.e. for every $\alpha>0$, $\rho(\alpha X)=\alpha \rho(X)$,
\item[v)] {\it sublinearity}, i.e. $\rho(X+Y)\leq \rho(X)+\rho(Y)$.
\end{itemize}
\end{definition}

Monotonicity represents the minimal requirement for a risk measure to model the preferences of a rational agent. If in addition conditions iii), iv) and v) hold then the risk measure is called coherent and it is automatically convex. Unfortunately both axioms iv) and v) appear to be restrictive and unrealistic: the former does not sense the presence of liquidity risks, the latter does not describe the real intuition hidden behind the diversification process. For this reason in most of the literature iv) and v) are substituted by ii) which has a natural interpretation: the risk of the diversified aggregated position $t X_1 + (1-t) X_2$ is surely smaller than the combination of the two single risks. Cash additivity is a key property which allows to characterize the risk procedure in terms of capital requirements
$$\rho(X)=\inf\{\alpha\mid X+\alpha \in \mathcal{A}\}$$
where $\mathcal{A}=\{X\in L^{\infty}\mid \rho(X)\leq 0\}$ is the acceptance sets. The risk of a position $X$ is thus the minimal amount of money that I have to save today in order to make the position acceptable with respect to a precise criterion (represented by the acceptance set $\mathcal{A}$) which is usually imposed by regulation agencies (External Risk Measures). On the other hand an institution may have some specific criteria that need to be enclosed in their model as in the case of Internal Risk Measures.
\bigskip

In literature, some extensions of the notion of risk measure have been considered in order to better describe the complexity of the risk process. We now recall the notion of set valued risk measures presented in \cite{JM04}. Given any subset $A\subset \R^d$ we shall denote by $L_d^p(A)$ the collection of $A$-valued random variable $X=(X_1,...,X_d)$ with finite $L^p$ norm (or equivalently $X_i\in L^p$ for every $i=1,...,d$). Whenever no confusion arises we denote $L^p_d=: L^p_d(\R)$. Notice that for $p=+\infty$ we end up with essentially bounded random vectors of dimension $d$.
\\
In this paper we take into account the theoretical framework developed in \cite{JM04} and \cite{HH10}: consider a closed convex cone $K^d\subsetneq \R^d$ (resp. $K^n\subsetneq \R^n$) such that $\R^d_+\subseteq K^d$ (resp. $\R^n_+\subseteq K^n$) and define the partial ordering $\preccurlyeq_d$  on $\R^d$ by $x\preccurlyeq_d 0$ iff $x\in K^d$ (similarly for $\preccurlyeq_n$ on $\R^n$). This ordering can be naturally extended to $L^{\infty}_d$ in the following way:
$$X\preccurlyeq Y\quad \Leftrightarrow \quad X-Y\in K^d \;\Prob\text{-almost surely}$$
Hence $L^{\infty}_d(K^d)$ is a cone that consists in all the non-negative random variables in the sense of $\preccurlyeq$.
\\Moreover for any $A,B\subseteq \R^n$ we may define the partial order $\preccurlyeq_n$ as
$$A\preccurlyeq_n B \quad \Leftrightarrow \quad B\subseteq A+K^n$$
We will indicate by $(1,1,...,1)=:\mathbf{1}_d \in \R^d$ ($\mathbf{1}_n \in \R^n$) the vector such that each component is equal to $1$, by $\mathbf{1}_d^j\in\R^d$ ($\mathbf{1}_d^j\in\R^d$) the vector such that the j$^{th}$ component is $1$ and the other are $0$.  Finally the sum among sets is to be intended as the usual Minkwoski sum.

\begin{definition}\label{RM} \emph{A $(d,n)$-risk measure is a set valued map $R:L^{\infty}_d\rightrightarrows \R^n$ satisfying the following axioms:}
\begin{description}
  \item{i)} for all $X\in L^{\infty}_d$, $R(X)$ is closed and $0\in R(0)\neq \R^n$;
  \item{ii)} for all $X,Y\in L^{\infty}_d$: $X\preccurlyeq_d Y$ $\Prob$-almost surely implies $R(Y)\preccurlyeq_n R(X)$.
\end{description}

In particular a $(d,n)$-risk measure is \emph{convex} if

\begin{description}
\item{iii)} for all $X_1,X_2\in L^{\infty}_d$, $\forall t\in [0,1]$,  $R(t X_1 + (1-t) X_2) \preccurlyeq_n t R(X_1) + (1-t) R(X_2)$
\end{description}
A $(d,n)$-risk measure is \emph{cash additive}  if
\begin{description}
\item{iv)} for all $x\in\R$, $j=1,...,d$ and $X\in L^{\infty}_d$ we have $R(X+x\mathbf{1}^j_d)=-x\mathbf{1}_n+R(X)$;
\end{description}

and \emph{coherent} if

\begin{description}
  \item{v)} it is sublinear: for all $X,Y\in L^{\infty}_d$, $R(X+Y)\preccurlyeq_n R(X)+R(Y) $ and
  \item{vi)} positively homogenous: for all $t>0$ and $X\in L^{\infty}_d$, $R(tX)=tR(X)$.
\end{description}
\end{definition}

\begin{remark}We now illustrate the financial meaning of the previous axioms.

\begin{enumerate} 

\item Monotonicity implies that for $X\preccurlyeq_d Y$ the efficient points of $R(X)$  \textquoteleft dominate \textquoteright  those of $R(Y)$. This domination might be described as follows: let $l$ be any straight line with origin in $0\in\R^n$ and passing through the positive orthant then the point $l\cap \partial R(X)$ is componentwise greater then $l\cap \partial R(Y)$. If in particular both $R(X)$ and $R(Y)$ admit an ideal point $x,y\in\R^n$ then $x$ is componentwise greater than $y$. This property describes exactly the idea that $X$ is riskier than $Y$ in terms of the geometry of the frontier of the sets $R(X)$ and $R(Y)$. A similar observation might be repeated for all the other properties: convexity/sublinearity state that the diversification/aggregation decrease the risk of the portfolio and positively homogeneity excludes any liquidity risk.

\item It is clear that if $R(X)+K^n=R(X)$ for every $X\in L^{\infty}_d$ the region which lies above the frontier $\partial R(X)$ represents higher level of risk that might be taken into account. This become a key point if we understand each component of $R(X)$  as different criteria. Consider for instance two agents endowed with two different risk measures $\rho_1$ and $\rho_2$: we define $R(X)=\mathcal{C}(\rho_1(X),\rho_2(X))$ where $\mathcal{C}(\rho_1(X),\rho_2(X))$ is a pointed cone in $\R^2$. We will be able to find an agreement between the agents only if $\mathcal{C}(\rho_1(X),\rho_2(X))\cap \{(x,y)\in\R^2 \mid x=y\}\neq \emptyset$.

\item Cash additivity in this multi-criteria framework slightly differs from the definition adopted in \cite{JM04}. The interpretation is straightforward: if we add to the $j^{th}$ component of the portfolio $X\in L^{\infty}_d$ a sure amount of money $x$ then all the different criteria will agree that the risk decreases by $-x$.  Every cash additive $(d,n)$-risk measure can be characterized by means of an acceptance set, namely a closed convex cone $\mathcal{A}\subset L^{\infty}_d$ and containing $L^{\infty}_d(K)$. In fact the set valued map defined as
$$R_{\mathcal{A}}(X)=\left\{\big(\sum_{j=1}^{d} x_j\big) \mathbf{1}_n \in\R^n\mid x=(x_1,...,x_d)\in \R^d \text{ and } X+x\in\mathcal{A}\right\}$$
is cash additive $(d,n)$-risk measure. Viceversa the acceptance set induced by a cash additive $R$ is given by
$$\mathcal{A}_{R}=\{X\in L^{\infty}_d\mid R(X)\preccurlyeq_n R(0) \}$$
\end{enumerate}
\end{remark}

\begin{example} Definition \ref{RM} show a simpler expression as soon as for every $X$ the set $R(X)$ satisfies $R(X)+K^n=R(X)$, as discussed in \cite{JM04} and \cite{HH10}. On the other hand in certain cases we may need to provide a confidence interval of risk so that we expect the risk measure to take values in compact sets. We illustrate this idea through an easy example: again let $1,2$ be two agents endowed with two different risk measures $\rho_1$ and $\rho_2$ which are estimating the risk of a position $X$. Suppose that $1$ only knows that $2$ is more conservative (i.e. $\rho_1(X)<\rho_2(X)$),  but does not know \emph{a priori} the procedure used by $2$. Thus agent $1$ will provide a confidence interval $[\rho_1(X),M]$, where $M$ represents the maximal capital requirement that $1$ is willing to hold in order to cover the risk of $X$. On the other hand agent $2$ will provide an interval $[m,\rho_2(X)]$ where $m$ is the minimal amount of money $2$ wants to save. Thus $R(X)=[\rho_1(X),M]\times [m,\rho_2(X)]$ describes the aggregate model: the agreement will take place only if $R(X)\cap \{(x,y)\in\R^2 \mid x=y\}\neq \emptyset$ which is equivalent to $[\rho_1(X),M]\cap [m,\rho_2(X)]\neq \emptyset$.

\end{example}

In the following Proposition we motivate the use of set valued risk measures instead of vector valued: if we consider different agents we are allowed to give different weights to each one of them, depending on the reliability.

\begin{proposition} let $l$ be any straight line with origin in $0\in\R^n$ and passing through the positive orthant $\R^n_+$ . Consider the map $\rho:L^{\infty}_d\rightarrow R^{n}$ defined as
\begin{equation}
\rho_l(X)=\inf \{R(X)\cap l\}
\end{equation}
where the $\inf$ is to be intended componentwise and $\rho_l(X)=+\infty\mathbf{1}_n$ if $R(X)\cap l=\emptyset$.
If $R$ is respectively monotone/convex/cash invariant/sublinear/positive homogeneous then $\rho$ is monotone/convex/cash invariant/sublinear/positive homogeneous.
\end{proposition}

Finally we state a well known automatic continuity result which is a key point for the optimization problems that we will consider in the core of this paper.

\begin{proposition}
\label{continuity}
\cite{JM04}
Every $(d,n)$-coherent risk measure $R:L^{\infty}_d\rightrightarrows \R^d$ such that $R(X)=R(X)+K^n$ for every $X\in L^{\infty}_d$, is continuous on $L^{\infty}_d$.
\end{proposition}

\section{Optimal Portfolio Diversification} \label{optimal}

Suppose that an agent is considering a vector $X=(X_1,...,X_d)\in L^{\infty}_d$ of risky financial positions. In a one-period model this vector might be composed by a basket of bonds, stocks, options so that every $X_i>0$ represents the value of the the $i^{th}$ position at the final time. This is the case of the two examples given in Section \ref{ModelAmb}. For sake of simplicity we assume that the price of each asset at time $0$ is equal to $\pi(X_i)=1$ for $i=1,...,d$. Here the pricing rule $\pi$ is given endogenously, in the sense that prices are fixed \emph{a priori} by market itself. The initial endowment $x\in\R$ will be given by a particular combination $\widehat{\alpha}\in\R^d_+$ so that $1=\sum_i\widehat{\alpha}_i \pi(X_i)=\sum_i\widehat{\alpha}_i$.
\\A different point (as the one followed in Section \ref{var}) is to consider a vector $(X_1,...,X_d)$ that describes the possible losses/gains that the decision maker (as an insurance company) has to face holding the position $X_i$.
\\
In both cases an admissible risk diversification strategy will be given by any vector $\alpha\in\R^d_+$ such that $\sum_i\alpha_i=1$, which represents the proportion of capital invested in each risky position.
\\The decision maker is interested in redistribute and minimize his risk, by means of an optimal strategy. Namely given a real valued risk measure (as introduced in \cite{ADEH}) $\rho:L^{\infty}\rightarrow \R$ and $X\in\R^d$ we have the following optimization problem
\begin{equation}\label{RealOpt}\min_{\alpha\in \Delta^d} \rho\left(\alpha\cdot X \right)\quad \text{ where } \quad \Delta^d=:\left\{\alpha\in\R^d_+ \mid \sum_i\alpha_i=1\right\}
\end{equation}

Here we extend the optimization problem (\ref{RealOpt}) to the case of set-valued risk measures giving a different interpretation than the one in \cite{JM04}, as explained in the following. Given a set-valued risk measures $R:L^{\infty}_d\rightrightarrows \R^n$  and $X\in L^{\infty}_d $,
the agent deals with the following set-valued optimization problem
\begin{equation}
\label{SetValuedOpt}\min_{\alpha\in \Delta^d} R\left(\alpha_1 X_1,..., \alpha_d X_d \right) \quad \text{ where } \quad \Delta^d=:\left\{\alpha\in\R^d_+ \mid \sum_i\alpha_i=1\right\}
\end{equation}
Since the vector $X\in\R^d$ is supposed to be fixed we will consider, with a slight abuse of notation, $R:\Delta^d\rightrightarrows \R^n$, so that we will often write $R(\alpha)$ instead of $R\left(\alpha_1 X_1,..., \alpha_d X_d \right)$.
\\We suppose $\R^n$ being ordered by the usual Pareto cone $\R^n_+$ which means $a\ge b$ if and only if $a-b\in\R^n_+$ for all $a,b\in\R^n$.
A pair $(\hat\alpha,\hat y)$, with $\hat y\in R(\hat\alpha)$ is an optimal solution
to (\ref{SetValuedOpt}) if $R(\alpha)\subseteq \hat y - (\R^n_+\backslash \{0\})^c$ for all $\alpha\in\Delta^d$ (see \cite{Aubin} for more details).
\\In the sequel we will consider the case in which the set $R(\alpha)$ takes the form
$$R\left(\alpha_1 X_1,..., \alpha_k X_k \right) := \check{R}\left(\alpha_1 X_1,..., \alpha_k X_k \right) + D(\alpha_1, \alpha_2, \dots, \alpha_n),$$
where $0 \in D(\alpha)\subseteq \R^n_+$ for all $\alpha\in\R^d_+$.

\begin{figure}[tb]
\begin{center}
\begin{tabular}{c}
\includegraphics[scale=0.6]{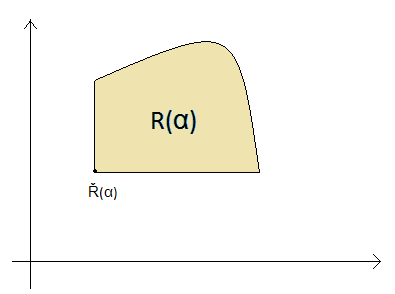}
\end{tabular}
\end{center}
\caption{A set-valued risk measure}
\label{setvalued}
\end{figure}

\noindent A particular case happens when $D=\R^n_+$ which leads to $R\left(\alpha\right)=\check{R}\left(\alpha\right)+\R^n_+$.
For instance when $n=1$, the above condition degenerates to $R\left(\alpha\right)=[\check{R}\left(\alpha\right),+\infty)$.
\\The vector $\check{R}$ can be thus seen as a conglomerate of these different risk attitudes: some may be coming from regulatory agencies  (external risk measures), others may describe different points of view arising inside the institution (internal risk measures).
Risk managers would like to choose a strategy that minimizes all these different points of view, but this clearly becomes a quite tough task.
\\Whenever the risk measure $R:L^{\infty}_d\rightrightarrows \R^n$ is convex then $R:\Delta^d \rightrightarrows \R^n$ inherits convexity for any fixed basket $X\in\R^d$. Thus the above program (\ref{SetValuedOpt}) admits a solution since $\Delta^d$ is a compact set. Moreover (\ref{SetValuedOpt}) can be solved by searching for solutions to the following problem (\ref{VectorOpt}):

\begin{equation}\label{VectorOpt}
\min_{\alpha\in \Delta^d} \check{R}\left(\alpha_1 X_1,..., \alpha_d X_d \right) \quad \text{ where } \quad \Delta^d=:\left\{\alpha\in\R^d_+ \mid \sum_i\alpha_i=1\right\}.
\end{equation}

In fact, it is easy to prove that solutions to (\ref{VectorOpt}) are actually solutions to (\ref{SetValuedOpt}).
Suppose that $\hat\alpha$ is an optimal solution to (\ref{VectorOpt}). Since $\hat\alpha$ is an optimal solution to (\ref{VectorOpt}) then it holds
\begin{equation}
\check{R}(\alpha)\subseteq \check{R}(\hat\alpha)- (\R^n_+\backslash\{0\})^c
\end{equation}
for all $\alpha\in\Delta^d$. By easy computations we get
\begin{equation}
R(\alpha) = \check{R}(\alpha) + D(\alpha) \subseteq \check{R}(\hat\alpha)-(\R^n_+\backslash\{0\})^c + D(\alpha) = \check{R}(\hat\alpha)-(\R^n_+\backslash\{0\})^c
\end{equation}
which shows that the pair $(\hat\alpha,\check{R}(\hat\alpha))$ solves (\ref{SetValuedOpt}). In the next section we propose a GP
model for finding approximate solutions to (\ref{VectorOpt}).

\section{Risk Management through a GP model with satisfaction function} \label{MOP}

In classical multi-criteria decision aid (MCDA) the agent has to consider several conflicting and incommensurable
objectives or attributes which have to be optimized simultaneously. If $D$ is the set of feasible solutions
and $f_i$ represents the $i$-th objective function then the general formulation of a MCDA model is as follows
\cite{Sawaragi}: Maximize  $(f_1(x),f_2(x),\ldots f_n(x))$ subject to the condition that $x\in D$.
We suppose that each $f_i$ is continuous and $D$ is a compact set which guarantee that
Weierstrass theorem applies providing the existence of a solution.
The Goal Programming model is a well known strategy for solving MCDA models; in this context, the agent seeks the best compromise
between the achievement levels $f_i(x)$ and the aspiration levels or goals $g_i$ by minimizing the absolute deviations
(see \cite{Aouni,Hannan,Larban,Martel,Romero}) and it can formulated as follows:
\begin{gather}
\min  \sum_{i=1}^n \delta_i^+ + \delta_i^-
\label{GP}\\
\text{s.t. }\left\{
\begin{array}[c]{l}
 f_i(x)+\delta_i^+-\delta_i^-=g_i, \ i=1\ldots n  \\
x\in D \\
\end{array}
\right.
\label{constrs}
\end{gather}
The GP model with weights (WGP), which represents an extension of (\ref{GP}), reads as:
\begin{gather}
\min  \sum_{i=1}^n w_i^+ \delta_i^+ + w_i^- \delta_i^-
\label{WGP}\\
\text{s.t. }\left\{
\begin{array}[c]{l}
 f_i(x)+\delta_i^+-\delta_i^-=g_i, \ i=1\ldots n  \\
x\in D \\
\end{array}
\right.
\label{wconstrs}
\end{gather}
where $w_i^+$ and $w_i^-$ are weights or scaling factors.
The GP model and its extensions have obtained a lot of popularity and attention because they represent
simple models to be analyzed and implemented. However it is worth underlying that an optimal solution to (\ref{GP}) and
(\ref{WGP}) is an optimal solution to (\ref{VectorOpt}) if some optimality tests are satisfied (\cite{Larban}).
Among several extensions of these models which are currently available, it is worth mentioning
the one developed by Martel and Aouni \cite{Martel} which explicitly incorporates the agent's preferences.
They introduced the concept of satisfaction function in the GP model where the agent can explicitly express his/her preferences
for any deviation between the achievement and the aspiration level of each objective.
In general, given three positive numbers $\xi_i,\xi_d$ and $\xi_v$ which will be called, respectively, the {\it indifference threshold},
the {\it dissatisfaction threshold} and the {\it veto threshold} in the sequel, a satisfaction function
is a map $F:[0,\xi_v]\to [0,1]$ which satisfies the following properties:
\begin{itemize}
\item $F(x)=1$, for all $x\in [0,\xi_i]$, where $\xi_i$ is the indifference threshold,
\item $F(x)=0$ for all $x\ge [\xi_d,\xi_v]$, where $\xi_d$ is the dissatisfaction threshold,
\item $F$ is continuous and descreasing.
\end{itemize}
Depending on the thresholds' values, which strictly depend on the agent preferences,
it could happen that positive and the negative deviations are penalized in a different manner.
The GP model with satisfaction function is formulated as follows:
\begin{gather}
\min  \sum_{i=1}^n w_i^+ F(\delta_i^+) + w_i^- F(\delta_i^-)
\label{GP1}\\
\text{s.t. }\left\{
\begin{array}[c]{l}
f_i(x)+\delta_i^+-\delta_i^-=g_i, \ i=1\ldots n  \\
x\in D \\
\delta_i^+,\delta_i^-\in [0,\xi_v]
\end{array}
\right.
\label{constrs}
\end{gather}
Let us notice that the GP model (\ref{GP1}) admits a solution because of the continuity of $f_i$ and $F$
and the compactness of $D$. Some applications of this model can also be found in \cite{Aouni1, Aouni2, Aouni3, Aouni4, Aouni5}.

\bigskip

Let us now formulate a GP model with satisfaction function for risk management and optimal portfolio diversification
based on the above multi-criteria optimization model (\ref{VectorOpt}):

\begin{gather}
\max  \sum_{i=1}^n w_i^+ F(\delta_i^+) + w_i^- F(\delta_i^-)
\label{GP2}\\
\text{s.t. }\left\{
\begin{array}[c]{l}
\check{R_i}\left(\alpha\right) +\delta_i^+-\delta_i^-=g_i, \ i=1\ldots n  \\
\sum_i\alpha_i=1 \\
\alpha_i\ge 0, \ i=1\ldots p \\
\delta_i^+,\delta_i^-\in [0,\xi_v]
\end{array}
\right.
\label{constrs}
\end{gather}

where $g_i$ represents the target of $\check{R_i}$ for all $i=1\ldots n$.
If $\hat\alpha$ solves (\ref{GP2}) and $\hat\alpha$ is an optimal solution to (\ref{VectorOpt}) then
$(\hat\alpha,\check{R}(\hat\alpha))$ solves (\ref{SetValuedOpt}). For this purpose one can use some
of the optimality tests presented in \cite{Larban}.

\section{Examples} \label{examples}

In this section we provide three possible applications of set valued risk measures to optimal risk diversification in presence of ambiguity concerning the risk criterion that should be adopted. In order to guarantee a simple computational procedure we choose a common mathematical framework for these examples, which is described in the following paragraph.
\\In the first and second example we suppose that the agent is uncertain on the probabilistic model he will have to choose: namely a set of possible probabilistic scenarios $\{\Q_i\}_{i=2}^{n+1}$ is taken into account. Notice that $\Q_1$ is excluded since it corresponds to the real probability $\Prob$.  Moreover we assume $\Q_i<<\Prob$ so that even though $\Prob$ is unknown we have that the null sets are fixed a priori. The standard approach would be to define the coherent risk measure $\rho(X)=\sup_{i=2,...,n+1}E_{\Q_i}[-X]$ and solve the optimization problem given by equation (\ref{RealOpt}). Here we will compare this deeply conservative approach with a multi objective goal programming method.
\\In the third example the reference probability is assumed to be known and we consider an agent that adopts the $V@R$ as a risk estimator. Since we are considering a problem of diversification over risky positions and the $V@R$ fails to be convex, the agent is uncertain on the criterium he will have to choose between the convex combination of the different Value At Risks $\sum_{i} \alpha_iV@R_{\lambda}(X_i)$ or the Value At Risk of the convex combination $ V@R_{\lambda}(\sum_{i}\alpha_iX_i)$.

\subsection{Robust methods for risk evaluation under model uncertainty.}\label{ModelAmb}

\paragraph{Illustrative setting} In all the following computational examples we fix  $$(\Omega,\F,\Prob)=([0,1],\mathcal{B}_{[0,1]},Leb)$$ and the sequence of functions
$$f_i(\omega)= \frac{i^{-\omega}}{\int_0^1 i^{-\omega} d\omega}\quad \forall\,i=2,...,n+1.$$
This last equation defines a sequence of probability $\Q_i$ such that $\frac{d\Q_i}{d\Prob}=f_i$ and $\Q_i\sim\Prob$.
We simply compute $\int_0^1 i^{-\omega} d\omega=\frac{1}{\ln i}\left(1-\frac{1}{i}\right)$ so that we deduce that $\lim_i f_i(0)=+\infty$ and $\lim_i f_i(\omega)=0$ for every $\omega>0$.

\paragraph{The linear case} We firstly consider an example in which the agent has different beliefs in terms of probabilistic models but no risk aversion. We consider a general portfolio $(\alpha_1 X_1,..., \alpha_d X_d)\in L^{\infty}_d$ where $\alpha\in\Delta^d$. We define the set valued map $R:L^{\infty}_d\rightrightarrows \R^n$ as
$$R(\alpha_1 X_1,..., \alpha_d X_d)=\prod_{i=2}^{n+1} \big[E_{\Q_i}\big[-\alpha\cdot X\big],+\infty\big),$$
where $\alpha\cdot X$ is the usual scalar product in $\R^d$. Simple computations show that $R$ is a $(d,n)$-risk measure satisfying (i), (ii), (iv), (v) and (vi) in Definition \ref{RM}.
\\In our illustrative setting we fix a portfolio composed by a non risky asset $X_1(\omega)=1$ and two risky assets $X_2(\omega)=2\omega$, $X_3(\omega)=3\omega^2$. The portfolio selection will be thus given by $\alpha_1,\alpha_2,\alpha_3\in [0,1]$ such that $\alpha_1+\alpha_2+\alpha_3=1$. We observe that
\begin{eqnarray*}E_{\Q_i}\left[\alpha\cdot X \right]&=&E_{\Q_i}\left[\alpha_1 X_1+\alpha_2 X_2+\alpha_3 X_3\right]
\\&=&\frac{1}{\int_0^1 i^{-\omega} d\omega}\int_0^1 (\alpha_1+2\alpha_2\omega+3\alpha_3\omega^2)i^{-\omega}d\omega
\\&=& \alpha_1-\frac{2\alpha_2}{\int_0^1 i^{-\omega} d\omega}\cdot\frac{1}{i\ln i}+\frac{2\alpha_2}{\ln i}-\frac{3\alpha_3}{\int_0^1 i^{-\omega} d\omega}\cdot\frac{1}{i\ln i}
\\&-& \frac{6\alpha_3}{\int_0^1 i^{-\omega} d\omega}\cdot\frac{1}{i(\ln i)^2}+\frac{3\alpha_3}{\ln i}
\\&=& \alpha_1+2\alpha_2\left(\frac{1}{\ln i}-\frac{1}{i-1}\right)+3\alpha_3\left(\frac{1}{\ln i}-\frac{1}{i-1}-\frac{2}{(i-1)\ln i}\right)
\end{eqnarray*}
Notice that the classical approach would suggest to compute
$$\rho_n\big( \alpha\cdot X\big)=\sup_{i=2,...,n+1} E_{\Q_i}\left[-\alpha\cdot X \right]=-\inf_{i=2,...,n+1} E_{\Q_i}\left[\alpha\cdot X \right]$$
Clearly $E_{\Q_i}\left[\alpha\cdot X \right]\geq 0$ and $E_{\Q_i}\left[\alpha\cdot X \right]\downarrow\alpha_1$ as $i\rightarrow +\infty$ so that
$$\rho_n\big( \alpha\cdot X \big)\stackrel{n\rightarrow \infty}{\longrightarrow} -\alpha_1 .$$
and the only strategy allowed becomes $\alpha_1=1$. This means that the higher is the number of probabilistic scenarios taken into account, the more the decision maker will concentrate his investments on the non risky asset.

We restrict ourselves to the case $i=2,3,4$. If we minimize $E_{\Q_i}\left[-\alpha\cdot X \right]$ separately  we obtain the three possible ideal goals $g_1=g_2=g_3=-1$ with respect to the three different criteria. These goals are always realized by the strategy $\alpha_1=1,\alpha_2=\alpha_3=0$. In this way no risk is hazarded and consequently no real gains are reached. On the other hand if we choose three goals $g_1,g_2,g_3$ which allow a slightly higher level of risk then a better performing strategy arises.
\\Let us choose the satisfaction function $F(\delta)={1\over 0.01*\delta^2}$, $\xi_v=60$,
and $w_1^+=w_1^-=w_2^+=w_2^-=w_3^+=w_3^-={1\over 3}$. We fix the goals $g_1$, $g_2$, and $g_3$ to be equal to $-0.7$,$-0.6$,$-0.5$.
The above GP model for optimal risk diversification can be formulated in this setting as follows:

\begin{gather}
\max \sum_{i=2}^4 \left( w_i^+ F(\delta_i^+) + w_i^- F(\delta_i^-)\right)
\label{GP_application}\\
\\ \text{s.t. }\left\{
\begin{array}[c]{l}
E_{Q_2}[-\alpha_1 X_1- \alpha_2 X_2 - \alpha_3 X_3]+\delta_2^+-\delta_2^-=g_2, \ \\
E_{Q_3}[-\alpha_1 X_1- \alpha_2 X_2 - \alpha_3 X_3]+\delta_3^+-\delta_3^-=g_3, \ \\
E_{Q_4}[-\alpha_1 X_1- \alpha_2 X_2 - \alpha_3 X_3]+\delta_4^+-\delta_4^-=g_4, \ \\
\alpha_1+\alpha_2+\alpha_3\le 1,\;
\alpha_1,\alpha_2,\alpha_3\ge 0\\
0 \le \delta_2^+,\delta_2^-,\delta_3^+,\delta_3^-,\delta_4^+,\delta_4^- \le \xi_v
\end{array}
\right.
\label{constrs}
\end{gather}

The numerical solution is performed by using LINGO 12 and provides the following solutions
$\alpha_1=0.000000$, $\alpha_2=0.9510359$, and $\alpha_3=0.4896409E-01$. The decision maker clearly prefers to invest in $X_2$ even though the probability $\Q_i(X_2\geq X_1)\leq \frac{1}{2}$ for every $i=2,3,4$.
\\With respect to the classical optimal portfolio diversification in which the agent is forced to invest only on the non-risky asset, here we obtain a less conservative optimal solution which minimizes the distance between each achievement level and its goal.

\paragraph{Entropic risk measure under ambiguity of the risk aversion.}
Again we suppose that the decision maker is uncertain on the probabilistic model $\Q_i<<\Prob$ and his preferences are described by some exponential utility $u(x)=1-e^{-\lambda x}$. Moreover the agent might be more confident about some probabilistic scenarios, so that his risk aversion $\lambda>0$ will depend on $\Q_i$ (i.e. $\lambda(\Q_i)=\lambda_i$). As usual the exponential utility  induces the entropic risk measures
$$\rho_{\Q_i}(X)=\inf\{m\in\R\mid E_{\Q_i}[u(X+m)]\geq u(0)\}=\frac{1}{\lambda_i}\ln E_{\Q_i}\left[e^{-\lambda_i X}\right].$$
We consider a portfolio $(\alpha_1 X_1,..., \alpha_d X_d)\in L^{\infty}_d$ where $\alpha\in\Delta^d$ and define a map $R:L^{\infty}_d\rightrightarrows \R^n$ as
$$R(\alpha_1 X_1,..., \alpha_d X_d)=\prod_{i=1}^k \big[\rho_{\Q_i}\big(\alpha \cdot X\big),+\infty\big)$$
Simple computations show that $R$ is a $(d,n)$ convex cash additive risk measure (i.e. satisfies (i), (ii), (iii) and (iv) in Definition \ref{RM}).
\\In our illustrative setting we fix a portfolio composed by a non risky asset $X_1(\omega)=1$ and two risky assets $X_2(\omega)=2\omega$, $X_3(\omega)=(4\omega-1)\mathbf{1}_{\{\omega\geq \frac{1}{4}\} }$. The portfolio selection will be thus given by $\alpha_1,\alpha_2,\alpha_3\in [0,1]$ such that $\alpha_1+\alpha_2+\alpha_3=1$. We thus have
\begin{equation*}\rho_{\Q_i}\big(\alpha\cdot X\big)=\frac{1}{\lambda_i}\ln\left( \frac{\int_0^1 \exp\{-\lambda_i(\alpha_1+2\alpha_2\omega+\alpha_3(4\omega-1)\mathbf{1}_{\{\omega\geq \frac{1}{4} \}})-\omega\ln i\}d\omega}{\int_0^1 i^{-\omega} d\omega}\right)
\end{equation*}
Computing explicitly the integrals one gets
\begin{eqnarray*}&&\rho_{\Q_i}\big(\alpha\cdot X\big)=\frac{1}{\lambda_i}\ln\left( i\ln i\right)-\frac{1}{\lambda_i}\ln\left( i-1\right)-\lambda_i\alpha_1+
\\&+&\ln\left(\frac{1-e^{-\frac{1}{4}(2\alpha_2\lambda_i+\ln i)}}{2\alpha_2\lambda_i+\ln i}+\frac{e^{-\frac{1}{4}(2\alpha_2\lambda_i+4\alpha_3\lambda_i+\ln i)}-e^{-(2\alpha_2\lambda_i+4\alpha_3\lambda_i+\ln i)}}{e^{\alpha_3}(2\alpha_2\lambda_i+4\alpha_3\lambda_i+\ln i)}\right)
\end{eqnarray*}
Again we solve the optimal portfolio diversification problem via a GP model. We restrict ourselves to the case $i=2,3,4$. If we minimize $\rho_{\Q_i}\left(\alpha\cdot X\right)$ separately  we obtain the three possible ideal goals $g_1=-4.405194, \,g_2=-5.497588, \,g_3=-6.575445$ with respect to the three different criteria.
\\As in the previous model, let us choose the satisfaction function $F(\delta)={1\over 0.01*\delta^2}$, $\xi_v=60$,
and the weights $w_1^+=w_1^-=w_2^+=w_2^-=w_3^+=w_3^-={1\over 3}$. We set the parameters $\lambda_2$, $\lambda_3$, and $\lambda_4$ equal to $4$, $5$ and $6$,
and the goals $g_1$, $g_2$, and $g_3$ equal to $-4$,$-5$, and $-6$ respectively.
The above GP model for optimal risk diversification can be formulated in this setting as follows:

\begin{gather}
\max \sum_{i=2}^4 \left( w_i^+ F(\delta_i^+) + w_i^- F(\delta_i^-)\right)
\label{GP_application}\\
\\ \text{s.t. }\left\{
\begin{array}[c]{l}
\rho_{Q_2}(\alpha_1 X_1+ \alpha_2 +X_2 + \alpha_3 X_3)+\delta_2^+-\delta_2^-=g_2, \ \\
\rho_{Q_3}(\alpha_1 X_1+ \alpha_2 +X_2 + \alpha_3 X_3)+\delta_3^+-\delta_3^-=g_3, \ \\
\rho_{Q_4}(\alpha_1 X_1+ \alpha_2 +X_2 + \alpha_3 X_3)+\delta_4^+-\delta_4^-=g_4, \ \\
\alpha_1+\alpha_2+\alpha_3\le 1,\;
\alpha_1,\alpha_2,\alpha_3\ge 0\\
0 \le \delta_2^+,\delta_2^-,\delta_3^+,\delta_3^-,\delta_4^+,\delta_4^- \le \xi_v
\end{array}
\right.
\label{constrs}
\end{gather}

The numerical solution is performed by using LINGO 12 and provides the following solutions
$\alpha_1=0.6659041$, $\alpha_2=0.000000$, and $\alpha_3=0.3340959$.

\subsection{A new point of view concerning $V@R$}\label{var}
In the previous examples we built up a set valued risk measure aggregating different real valued risk measures that were generated different probability beliefs. Moreover in both example $R$ depended on the vector $X=(X_1,...,X_d)\in L^{\infty}_d$ only through the sum of the components. In this last example we suppose that the historical probability measure is known: in this framework the most popular (and also most debated) risk measure is $V@R_{\lambda}$ defined as
$$V@R_{\lambda}(Y)=-\sup\{m\in\R\mid \Prob(Y\leq m)\leq \lambda\},$$
where $\lambda\in [0,01,0.05]$ and $Y\in L^0$ is any $\F$-measurable random variable.
Notice that for every $a>0$ we have $V@R_{\lambda}(a Y)=a V@R_{\lambda}(Y)$ and $V@R_{\lambda}( Y+c)=V@R_{\lambda}(Y)-c$ for every $c\in\R$. Nevertheless the Value at Risk is not convex on the space of random variables and for this reason it does not sense the effect of diversification. This lack has an immediate consequence:
if we consider a basket of financial instruments $X=(X_1,..., X_d)\in L^{\infty}_d$ and $\alpha\in\Delta^d$ we cannot guarantee any order relation between
\begin{equation} V_{\lambda}(\alpha)=:\sum_{i} \alpha_iV@R_{\lambda}(X_i)\quad \text{ and } \quad V^{\lambda}(\alpha)=:V@R_{\lambda}\left(\sum_{i}\alpha_iX_i\right)
\end{equation}
As a consequence the decision maker should be uncertain among all the possible values $x\in [V_{\lambda}(\alpha)\wedge V^{\lambda}(\alpha),V_{\lambda}(\alpha)\vee V^{\lambda}(\alpha)]$. This problem can be clearly reinterpreted via a multi objective goal programming.
\\Another common risk measure is given by $\rho_{w}(Y)=-ess\inf Y$ with $Y\in L^{\infty}$, which is a coherent (and thus convex) risk measure. The clear drawback is that worst case risk measure is too restrictive and conservative from the point of view of an agent who is investing his capital. Anyway it can be exploited to give an upper boundary of the maximal capital requirement necessary to cover any possible expected loss.
\\We introduce the following set valued map $R:L^{\infty}_d \rightrightarrows \R^2$ defined as
\begin{equation} R(X_1,...,X_d)=\left\{(x,y)\in\R^2: \begin{array}{c} V_{\lambda}(\mathbf{1}_d)\leq x \leq \rho_{w}(\mathbf{1}_d)
\\\text{ and }
\\ V^{\lambda}(\mathbf{1}_d)\leq y\leq  V^{\lambda}(\mathbf{1}_d)+\frac{\rho_w(\mathbf{1}_d)-V^{\lambda}(\mathbf{1}_d)}{\rho_w(\mathbf{1}_d)-V_{\lambda}(\mathbf{1}_d)}(x-V_{\lambda}(\mathbf{1}_d))  \end{array} \right\}
\end{equation}
where $V_{\lambda}(\mathbf{1}_d)=:\sum_{i} V@R_{\lambda}(X_i)$, $V^{\lambda}(\mathbf{1}_d)=:V@R_{\lambda}\left(\sum_{i}X_i\right)$ and $\rho_w(\mathbf{1}_d)=\sum_{i} \rho_w(X_i)$. Notice that the set $R(X_1,...,X_d)$ is a triangle and has a minimizer and a maximizer (w.r.t the Pareto cone $\R^2_+$) given respectively by $(V_{\lambda},V^{\lambda})$ and $(\rho_w,\rho_w)$.
\\We consider a vector $(\alpha_1 X_1,..., \alpha_d X_d)\in L^{\infty}_d$ where $\alpha\in\Delta^d$ so that by the positive homogeneity of $V@R$ and $\rho_w$ we find
\begin{equation*} R(\alpha_1X_1,...,\alpha_dX_d)=\left\{(x,y)\in\R^2: \begin{array}{c}  V_{\lambda}(\alpha) \leq x \leq \rho_w(\alpha)
\\\text{ and }
\\  V^{\lambda}(\alpha)\leq y\leq   V^{\lambda}(\alpha)+\frac{\rho_w(\alpha)-V^{\lambda}(\alpha)}{\rho_w(\alpha)- V_{\lambda}(\alpha)}(x-V_{\lambda}(\alpha))  \end{array} \right\}
\end{equation*}
where $\rho_w(\alpha)=\sum_{i} \alpha_i\rho_w(X_i)$.
\\As usual our illustrative setting is given by $(\Omega,\F,\Prob)=([0,1],\mathcal{B}_{[0,1]},Leb)$ but in this case the Lebesgue measure in chosen as the reference probability. In order to clarify the example we consider three financial positions which allow negative losses namely $X_1(\omega)=0$, $X_2(\omega)=2\omega-1$ and $X_{3}(\omega)=2-4\omega^2$. Since the three random variables are continuous we deduce that for $\lambda=0.05$
\begin{eqnarray*} V@R_{\lambda}(X_1)=0 & V@R_{\lambda}(X_2)=0,9 & V@R_{\lambda}(X_3)=1,61
\\\rho_w(X_1)=0 & \rho_w(X_2)=1 & \rho_w(X_3)=2
\end{eqnarray*}
so that $V_{\lambda}(\alpha)=\alpha_2\cdot0,9+\alpha_3\cdot1,61$ and $\rho_w(\alpha)=\alpha_2\cdot1+\alpha_3\cdot 2$. We need to compute $V^{\lambda}(\alpha)$: notice that $\sum_i\alpha_i X_i(\omega)= \alpha_2(2\omega-1)+\alpha_{3}(2-4\omega^2)$. In particular $\sum_i\alpha_i X_i(0)=2\alpha_3-\alpha_2=-\sum_i\alpha_i X_i(1)$. Then the function $\sum_i\alpha_i X_i(\omega)$ have a maximum point in $\omega= \frac{\alpha_2}{4\alpha_3}$. In general the minimum of the parabola $f(\omega)=\alpha_2(2\omega-1)+\alpha_{3}(2-4\omega^2)$ will fall on $\omega=0$ if $2\alpha_3-\alpha_2> 0$ and on $\omega=1$ if $2\alpha_3-\alpha_2< 0$.
\begin{itemize}
\item[(1)] Suppose that $\alpha_2=2\alpha_3$. In this case the $V@R$ can be simply computed as
$$V@R_{\lambda}\left(\alpha\cdot X\right)=f\left(\frac{\lambda}{2}\right).$$
\item[(2)] Assume $\alpha_2<2\alpha_3$. We find that the solution of $f(0)=f(\omega)$ is given by $\omega=0; \frac{\alpha_2}{2\alpha_3}$ and $\Prob\left(\alpha\cdot X\leq f(0)\right)=1-\frac{\alpha_2}{2\alpha_3}$. We have two possible cases
\begin{itemize}
\item[(a)] $\lambda \leq 1-\frac{\alpha_2}{2\alpha_3}$ then
$$V@R_{\lambda}\left(\alpha\cdot X\right)=f\left(1-\lambda\right).$$
\item[(b)] $\lambda > 1-\frac{\alpha_2}{2\alpha_3}$ then
$$V@R_{\lambda}\left(\alpha\cdot X\right)=f\left(\frac{\lambda-(1-\frac{\alpha_2}{2\alpha_3})}{2}\right).$$
\end{itemize}
\item[(3)] Assume $\alpha_2>2\alpha_3$. We find that the solution of $f(1)=f(\omega)$ is given by $\omega=1; \frac{\alpha_2}{2\alpha_3}-1$ and $\Prob\left(\alpha\cdot X\leq f(1)\right)=\frac{\alpha_2}{2\alpha_3}-1$. We have two possible cases
\begin{itemize}
\item[(a)] $\lambda \leq \frac{\alpha_2}{2\alpha_3}-1$ then
$$V@R_{\lambda}\left(\alpha\cdot X\right)=f\left(\lambda\right).$$
\item[(b)] $\lambda > \frac{\alpha_2}{2\alpha_3}-1$ then
$$V@R_{\lambda}\left(\alpha\cdot X\right)=f\left(1-\frac{\lambda-(\frac{\alpha_2}{2\alpha_3}-1)}{2}\right).$$
\end{itemize}

\end{itemize}
Finally, fixing $\lambda=0.05$ we can conclude
\begin{eqnarray*}\text{ if }  \alpha_2\leq 1.9\alpha_3 &\text{then}& V@R_{\lambda}\left(\alpha\cdot X\right)= 0.9\alpha_2-1,61\alpha_3
\\\text{ if }  1.9\alpha_3< \alpha_2\leq 2\alpha_3 &\text{then}& V@R_{\lambda}\left(\alpha\cdot X\right)= -4\alpha_3 \left(\frac{\alpha_2}{4\alpha_3}-0.475\right)^2+\frac{\alpha_2^2}{2\alpha_3}+2\alpha_3-1.95\alpha_2
\\\text{ if }  2\alpha_3< \alpha_2<2.1\alpha_3 &\text{then}& V@R_{\lambda}\left(\alpha\cdot X\right)= -0.9\alpha_2 +2.99\alpha_3
\\\text{ if }  2.1\alpha_3\leq \alpha_2 &\text{then}& V@R_{\lambda}\left(\alpha\cdot X\right)= -4\alpha_3 \left(\frac{\alpha_2}{4\alpha_3}+0.475\right)^2+\frac{\alpha_2^2}{2\alpha_3}+2\alpha_3-0.05\alpha_2
\end{eqnarray*}
The above GP model for optimal risk sharing can be formulated in this setting as follows:
\begin{gather}
\max \sum_{i=1}^2 \left( w_i^+ F(\delta_i^+) + w_i^- F(\delta_i^-)\right)
\label{GP_application}\\
\\ \text{s.t. }\left\{
\begin{array}[c]{l}
\sum_{i=1}^3 \alpha_i V@R^{\lambda}(X_i)+\delta_1^+-\delta_1^-=g_1, \ \\
V@R_{\lambda}(\sum_{i=1}^3 \alpha_i X_i)+\delta_2^+-\delta_2^-=g_2, \ \\
\alpha_1+\alpha_2+\alpha_3\le 1,\;
\alpha_1,\alpha_2,\alpha_3\ge 0\\
0 \le \delta_1^+,\delta_1^-,\delta_2^+,\delta_2^- \le \xi_v
\end{array}
\right.
\label{constrs}
\end{gather}

We choose the goals $g_1$ and $g_2$ to be equal to $0.3$ and $g_2 = 0.5$ respectively.
LINGO 12 provides the following optimal solution: $\alpha_1=0.6823770$, $\alpha_2=0.2151639$, and $\alpha_3=0.1024590$.

\section{Conclusions and further developments}

The recent notion of set-valued risk measure appears as powerful tool that can be exploited to overcome many complications that arise in risk management. Risk, understood as capital requirements needed to cover expected future losses, becomes an ambiguous factor to determine as far as a manager has to face different criteria or is uncertain on the real probabilistic model that lays underneath the financial problem. Inspired from the statistical notion of confidence intervals, the general definition presented in this paper allows to consider compact valued risk measures.  In this way we associate to any financial position a cloud of different risk levels, instead of a single number, taking into account all the multiplicity of ingredients that characterize this computation.  As illustrated by some examples  we may thus formulate an optimal risk diversification problem which allocates the risk of a given portfolio in an optimal manner. This is a set valued program which can be reduced to a vector valued model if the images of the set valued mapping admit an ideal point. Using a Goal Programming approach with satisfaction function we are able to provide  approximate solutions to this vector model: the presence of several free parameters is the strength of this approach since this allows a calibration of the model sensitive to the risk aversion of the agent.
\\
We have then illustrated three different examples which support this approach: in the first and the second one,  the agent has fixed a risk procedure $\rho$ but he is uncertain about the probabilistic model $\Q$. In such a case the functional form of $\rho_{\Q}(\cdot)$ will explicitly depend on $\Q$ and the standard literature would suggest to take a supremum $\sup_{\Q} \rho_{\Q}(\cdot)$ to compute the capital requirement. As explained above, such a strategy would often force the agent to avoid any risk in his decision. Through the GP model we find a non trivial diversification strategy which takes into account all these different possible scenarios $\Q$. In the third example we provide a case of compact-valued risk measure built up from the celebrated Value at Risk. As well known, the Valued at Risk is convex only on the space of Gaussian random variables, but it looses this property if we extend its domain to more general random variables. As a consequence the $V@R$ is not sensitive to diversification and for this reason it might not fit our optimization problem. The method we have proposed in an illustrative setting is a natural starting point to overcome this controversial and debated feature of the $V@R$.
\\
For future developments, we are going to conduct a statistical analysis of the model illustrated in this manuscript by using real data and by estimating the images of a set-valued risk measure through the analysis of confidence intervals.

\end{document}